\documentclass[manuscript, nonacm]{acmart}
\AtBeginDocument{%
  \providecommand\BibTeX{{%
    \normalfont B\kern-0.5em{\scshape i\kern-0.25em b}\kern-0.8em\TeX}}}

\setcopyright{acmcopyright}
\copyrightyear{2023}
\acmYear{2023}
\acmDOI{XXXXXXX.XXXXXXX}

\acmConference[The 1st International Workshop on Explainable AI for the Arts]{ ACM Creativity \& Cognition}
{June 19-21, 2023}
  
\begin{document}

\title{Human-Machine Co-Creativity with Older Adults -- A Learning Community to Study Explainable Dialogues}
\renewcommand{\shorttitle}{The 1st International Workshop on Explainable AI for the Arts, ACM Creativity and Cognition 2023}

\author{Marianne Bossema}
\email{m.bossema@hva.nl}
\orcid{0000-0002-0622-0552}
\affiliation{%
  \institution{Amsterdam University of Applied Sciences, Leiden University}
  \country{The Netherlands}
}

\author{Rob Saunders}
\affiliation{%
  \institution{Leiden University}
  \country{The Netherlands}
}

\author{Somaya Ben Allouch}
\affiliation{%
  \institution{Amsterdam University of Applied Sciences, University of Amsterdam}
  \country{The Netherlands}
}

\renewcommand{\shortauthors}{The 1st International Workshop on Explainable AI for the Arts, ACM Creativity and Cognition 2023}

\begin{abstract}
This position paper is part of a long-term research project on human-machine co-creativity with older adults. The goal is to investigate how robots and AI-generated content can contribute to older adults' creative experiences, with a focus on collaborative drawing and painting. The research has recently started, and current activities are centred around literature studies, interviews with seniors and artists, and developing initial prototypes. In addition, a course ``Drawing with Robots'', is being developed to establish collaboration between human and machine learners: older adults, artists, students, researchers, and artificial agents. We present this course as a learning community and as an opportunity for studying how explainable AI and creative dialogues can be intertwined in human-machine co-creativity with older adults. 
\end{abstract}



\received{2 May 2023}
\received[revised]{25 May 2023}

\maketitle
\textbf{Reference: \\} 
Marianne Bossema, Rob Saunders, and Somaya Ben Allouch. 2023. In the 1st International Workshop on Explainable AI for the Arts (XAIxArts), ACM Creativity and Cognition (C\&C) 2023. Online, 3 pages.

\section{Introduction}
The global population is rapidly aging. According to the United Nations \cite{united2022world}, the number of people aged 60 years or older is expected to more than double by 2050, reaching approximately 2.1 billion. This has significant social, economic, and health implications. Based on a review of 900 studies, the World Health Organization concluded that engaging in creative activities can promote health and well-being, and help prevent and slow age-related physical and cognitive decline \cite{fancourt2019evidence}. 
Social robots offer unique opportunities to support creativity through assistance and social interaction. In addition, advancements in generative AI bring new possibilities for creative collaborations between humans and machines. There are many open questions, however, about how to design human-machine co-creativity (HMCC) for and with the target group of older adults.

When studying the design of technology, there are several motivators for involving users throughout the research \cite{fischer2020importance}. In general, this allows for gaining insights into user needs and perspectives, and for collecting feedback on ideas and prototypes. In our case, it helps to derive grounded user-centered design principles for HMCC with older adults. It can also contribute to their empowerment by eliciting curiosity and positive feelings, increasing familiarity and trust with technology, and enriching creative experiences.

HMCC systems are ``mixed-initiative creative interfaces'', involving humans and computers in an interactive loop where each produces, evaluates, modifies, and selects creative outputs in response to the other \cite{deterding2017mixed}. An example is ``Drawing Apprentice'', a co-creative system that collaborates with users in real-time abstract drawing. HMCC emerges through dialogic interaction across multiple modalities including language, drawing, music, etc. Ethical considerations, such as transparency, bias, and power relations, must be taken into account from the beginning of the design research process, when investigating these dialogues.

\section{Human-centered XAI for co-creativity}
Current methods for XAI may not always offer explanations that are easily understandable to humans \cite{angelov2021explainable}. Instead, they often provide only superficial insights into the ``black box'' of AI models, through post hoc hints about specific features or local information. This approach differs significantly from human reasoning, which involves holistic comparisons of items, rather than focusing on individual features \cite{angelov2021explainable}. Visualizations can contribute to XAI. For example, saliency maps highlight the important features or inputs that contributed to a particular decision, helping users understand the decision-making process of an AI system for that specific case \cite{samek2017explainable}. Interactive visualizations allow users to manipulate the inputs or parameters of the model and visualize the resulting changes, helping them to gain insights into what can be expected, and build trust. The effectiveness of explanation, however, lies in the perception of the person receiving the explanation \cite{liao2021human}. In HMCC with visual artifacts, both visualizations and interactivity can contribute to XAI, and can be designed to meet the needs for explainability of non-experts, such as older adults. 

In creative collaborations, dialogues are essential to develop understanding and common ground, either language-based or through creative artifacts. A dialogic approach for computational creativity, as suggested by Bown et al. \cite{bown2020speculative}, enables both human and artificial agents to actively influence the creative process and products, and adapt to the other's behavior. HMCC studies with robots sometimes have used speech, e.g. for scaffolding creative tasks, and prompting creative reflection \cite{ali2021social,hubbard_child-robot_2021}. Both interactive visualisations and language-based (speech) dialogues may contribute to explainable HMCC for the target group of older adults. 

Llano et al. \cite{llano2022explainable} propose Explainable Computational Creativity (XCC) to develop systems that can effectively communicate and explain processes, decisions, and ideas throughout the creative process. They suggest two-way communication channels that promote co-creation, understandable to both humans and machines. For example, an artificial agent engaged in collaborative drawing, may ask questions about a human-made sketch and use the answers to adapt and make new suggestions, while the human interaction partner may interact with, and select suggested artifacts and thereby manipulate the parameters of the model. This type of bi-directional dialogue can promote reflexivity as being an important part of XAI and at the same time contribute to the creative process.

\section{A learning community exploring dialogues for explainable co-creativity}
We present a ``Drawing with Robots'' course for older adults as a learning community. The goal is to support a mutual, collaborative learning process for both human and artificial agents. In human-machine dialogues for co-creativity, elements of XAI can be explored, implemented, evaluated, and improved. Research in psychology showed that when children view a task as a performance situation, they tend to be more risk-averse and less persistent than children who view the same task as a learning situation. Those with a learning frame engage in more experimentation and are more likely to try out new strategies \cite{edmondson2003framing}. This framing of a learning situation may also be beneficial for older adults. Research has shown that compared to younger people, seniors are more likely to avoid trial-and-error strategies, and rely on existing knowledge\cite{sakaki2018curiosity, Romero2012CreativityIC}. This depends on conditions, however, such as the efforts required to set goals, take action, and evaluate results \cite{oudeyer2016intrinsic}. The setting of a learning community and the framing of an artificial agent as one of the learners can engage older adults to adopt new strategies, ask for explanations, and share perspectives.
 
In this learning community, agile design and development processes can help to improve the deployed technology iteratively. Older adults can develop an understanding of technological possibilities, allowing them to better think along and provide feedback. This will also offer insights into shifting needs for explainability, when users become more familiar with the technology over time. Overall, this project will contribute to a deeper understanding of how dialogues, either language-based or through visual artifacts, can support explainable HMCC with older adults. It also informs how to implement XAI strategies for this often overlooked target group. This will allow the research field to gain further insights in how dialogues can be designed to support XAI in creative processes.

\begin{acks}
This publication is part of the project `Social robotics and generative AI to support and enhance creative experiences for older adults', with project number 023.019.021 of the research program Doctoral Grant for Teachers financed by the Dutch Research Council (NWO).
\end{acks}

\bibliographystyle{ACM-Reference-Format}
\bibliography{Articles}

\end{document}